\documentclass[aps,prx,twocolumn,superscriptaddress]{revtex4-1}

\usepackage{graphicx}
\usepackage{dcolumn}
\usepackage{amsfonts,amsmath,amssymb,bm}
\usepackage{subcaption}
\usepackage{xcolor}
\usepackage[utf8]{inputenc}
\usepackage[colorlinks=true,allcolors=blue]{hyperref}

\begin{document}
\title{Tunable Ohmic contact in graphene/HfS$_2$ van der Waals heterostructure}
\date{\today}
\author{S. Karbasizadeh}
\affiliation{Department of Physics, Isfahan University of Technology, Isfahan, 84156-83111, Iran}

\author{F. Fanaeiparvar}
\affiliation{Department of Physics, Isfahan University of Technology, Isfahan, 84156-83111, Iran}

\author{I. Abdolhosseini~Sarsari}
\affiliation{Department of Physics, Isfahan University of Technology, Isfahan, 84156-83111, Iran}

\begin{abstract}
With the use of density functional theory calculations and addition of van der Waals correction, the graphene/HfS$_2$ heterojunction is constructed, and its electronic properties are examined thoroughly. This interface is determined as $n$-type Ohmic and the impacts of different amounts of interlayer distance and strain on the contact are shown using Schottky barrier height and electron injection efficiency. Dipole moment and workfunction of the interface are also altered when subjected to change in these two categories. The transition between Ohmic to Schottky contact is also depicted to be possible by applying a perpendicular electric field, proving this to be yet another useful method for tuning different properties of this structure. The conclusions given in this paper can exert an immense amount of influence on the development of two-dimensional HfS$_2$ based devices in the future.

\end{abstract}

\maketitle

\section{INTRODUCTION}
The discovery of graphene in the year 2004 \cite{Novoselov666} may very well have been the key to opening the eyes of humanity to a world of two-dimensional materials with features like no other possessed before. At the forefront of this phenomenal revolution stands graphene, a two-dimensional substance with superior properties such as extreme mechanical strength \cite{Lee385}, high optical transmittance \cite{Nair1308} and ultrahigh mobility of charge carriers at room temperature \cite{BOLOTIN2008351}. These characteristics have made this one of a kind material pretty suitable for use in catalysts \cite{C2EE22238H}, sensors \cite{C3TA11774J} and nanoelectronic devices \cite{6518205}. Still, there is one major flaw in the structure of graphene that keeps it from being the ideal ultimate two-dimensional body. It is well-known that graphene is a semiconductor with a band-gap value of zero, band-gap and this unfortunate trait has limited its applications in high-speed electronic devices. 

Graphene, however, is not the only two-dimensional mass that encounters this problem and many other structures have been rendered inefficient or unable to reach their full potential due to the issue of zero or small band-gaps. Thus, the focus of many scientists and researchers has been aimed at these substances, trying to find the best possible solution to open a wider band-gap and extend the applicability of said materials. Different approaches have been applied to this task; some of which include mechanical strains \cite{Johari2012}, substitutional doping \cite{Wang2012} and layer stacking \cite{PhysRevB.92.075439}. Of the three techniques, stacking of layers has proven to be the most valuable. An assembly of layers based off of two-dimensional matter is of utmost significance mostly since each substance is free of out-of-plane dangling bonds; resulting in a relatively weak van der Waals (vdW) interaction and weakened Fermi level pinning at interfaces \cite{Liue1600069}. 

These vertical heterostructures can help bring the next generation of nanoelectronic transistors into the light.  One example of a genuinely two-dimensional nano transistor constructed using heterostructures is one built out of layers of graphene, MoS$_2$ and hexagonal boron nitride~\cite{Roy2014}. In this nano transistor, graphene acts as both source or drain and gate electrodes, and MoS$_2$ is the channel whereas h-BN is the high-k dielectric.

In the epitaxial layers mentioned, graphene is used as the contact in the device rather than the channel. The same strategy has been successfully applied to several other systems, and many other ultrathin vdW heterostructures have been fabricated experimentally and investigated theoretically. The graphene has been put on top of other two-dimensional materials acting as the contact. To name a few of these two-dimensional bodies, the following can be mentioned: phosphorene~\cite{PhysRevLett.114.066803}, blue phosphorene~\cite{C7CP01852E}, GaSe~\cite{C8CP02190B}, GaS~\cite{Pham2018,mosaferi2018direct}, g-GaN~\cite{doi:10.1063/1.4982690}, SiC~\cite{C8CP03508C}, arsenene~ \cite{doi:10.1063/1.4935602}, graphene-like GeC~\cite{Wang2019} and ZrS$_2$~\cite{WANG2019778}. The graphene layer used as the contact easily outperforms traditional metallic substances, the cause of which is its superior qualities mentioned above. Graphene-based Schottky contact is, in particular, unlike the conventional metal-semiconductor interfaces (namely the silicon-metal interface). In these conventional devices, the Schottky barrier does not change with the work function of the metal, and the existence of surface states pins the Fermi level. The same cannot be said for the interface of graphene and a semiconductor, in which Fermi level pinning can be overcome~\cite{Sinha2014,PhysRevB.97.195447}. The graphene-based heterostructures frequently preserve the unique Dirac cone which provides a higher electronic quality for devices built out of them. 

After the discovery of graphene, transition metal dichalcogenides (TMDs) made an entrance on to the stage. This class of materials consists of one transition metal from the periodic table and two chalcogens attached to it; hence the name, transition metal dichalcogenide. Since the discovery of direct band gaps~\cite{Splendiani2010} in these two-dimensional bodies and their potential applications in electronics~\cite{Briggs_2019}, the field of TMD monolayers has been one of huge allure to the men of science and industry alike. 
The combination of TMDs and other two-dimensional materials  (van der Waals heterostructures)  such as graphene has found a substantial footing as building blocks in a range of remarkable devices including transistors, solar cells, LEDs, fuel cells and sensing instruments~\cite{Petoukhoff2016,Janisch_2016}. 
TMD monolayers are much more desirable in transistor applications than graphene; the reason being the substance's zero bandgaps resulting in a low on/off ratio. The stability in their structure, the existence of a bandgap and showing electron mobilities comparable to that of silicon makes these structures suitable for the job. The monolayers of TMD have been found to have lower electron mobility than their bulk counterparts; as having just one layer makes them more susceptible to damage. The coating of these monolayers with high-$k$ dielectrics such as hexagonal boron nitride~\cite{Manzeli2017},  can solve this problem.

The most thoroughly examined group among transition metal dichalcogenides is the VIB TMDs like molybdenum disulphide or MoS$_2$ with a 2H structure \cite{PhysRevLett.108.196802}. 
The IVB TMDs with 1T structures were said to possess better overall qualities than the VIB TMDs \cite{Zhang2014}; but despite this assessment, the amount of work done on the group remains limited. Hafnium disulphide (HfS$_2$) is a member of this group of TMDs. With properties that are very much analogous to zirconium disulphide (ZrS$_2$) \cite{doi:10.1002/pssb.201700033}, a reported bandgap of 2.02 eV obtained via GW functional \cite{doi:10.1002/pssb.201700033} and calculated room-temperature mobility of about five times bigger than that of MoS$_2$ \cite{Zhang2014}, HfS$_2$ is without a doubt a most promising candidate to replace silicon in electronic devices. This two-dimensional material becomes even more technologically appealing with the existence of high-$k$ native dielectric of HfO$_2$ \cite{Mleczkoe1700481}, which can be very instrumental in the construction of nano transistors. 
Few-layered HfS$_2$ has already been fabricated and employed to create field-effect transistors using HfO$_2$ as dielectric \cite{doi:10.1002/smll.201600521} and phototransistors exhibiting a high photosensitivity \cite{doi:10.1002/adma.201503864}. This work of fabrication, however, was performed based on micromechanical cleavage from single crystals, a method only applicable at a nanoscale, not for a large area deposition and not able to exfoliate one single layer for further use. Other more recent synthesis techniques have been at work and have accomplished the task of breaking up the layers, providing two-dimensional HfS$_2$ in high-yield \cite{Wang:2018:1533-4880:7319}. Chemical vapour deposition \cite{Wang2018} and liquid exfoliation \cite{Kaur2018} are two of the more dependable of these means. Growing HfS$_2$ films at low temperature (100 $^{\circ}$C) using atomic layer deposition (ALD) has also been made possible \cite{Cao2020}. The layers produced in this method have an adjustable thickness ranging from angstroms to $>$100 nm and are available on a large scale. ALD reaction is also in possession of a mixture of advantageous qualities when it comes to integrating this semiconductor into different devices. These procedures have done a fantastic job of opening the arena for further examination of HfS$_2$ fabric.    

In this paper, an ultrathin van der Waals heterostructure of graphene and HfS$_2$ monolayer referred to as G/HfS$_2$ throughout the rest of the article is designed and its structural and electronic properties are studied. The effects of interlayer distance and strain on this heterojunction are explored through the measurement of Schottky barrier height (SBH), orbital overlap, tunneling probability ($T_B$) and charge transfer. An external electric field is also applied, and its influence on inducing a transition from Ohmic to Schottky contact is determined. 

The remaining parts of the letter are arranged in this order: Sec. \ref{II} illuminates the computational means by which the calculations are performed. Sec. \ref{III} gives a full report of the results procured, dissecting them to the best of author's knowledge, while Sec. \ref{IV} concludes the paragraph in an orderly fashion.

\section{COMPUTATIONAL METHODOLOGY}\label{II}
The work carried out based on first-principles calculations residing within the framework of density functional theory (DFT) formalism \cite{PhysRev.140.A1133,PhysRev.136.B864} is conducted using Quantum Espresso simulation package \cite{Giannozzi_2009}. The projector augmented wave (PAW) \cite{PhysRevB.50.17953,PhysRevB.59.1758} pseudopotentials are utilized taking advantage of Perdew-Burk-Ernzerhof (PBE) generalized gradient approximation (GGA) \cite{PhysRevLett.77.3865} as the exchange-correlation functional. An electronic device where the semiconductor acts as the channel, the doping of it by the gate voltage or the metal electrode is inevitable. The result of this occurrence is the great screening of electron-electron interaction, making the single electron theory a compelling approximation in determining the band edge positions \cite{C5NR06204G}. Some previous studies have also conquered that the transport bandgap energy of the channel foreseen by GGA is much closer to the experimental amounts in comparison with the results obtained via the GW estimate \cite{Ataca2012,PhysRevB.88.195420}.

To give an accurate description of the long-range interactions between the two layers, Grimme's DFT-D2 \cite{doi:10.1002/jcc.20495,doi:10.1002/jcc.21112} van der Waals correction is put to work. Geometrical relaxation is executed on all the structures given until all the forces on each atom are smaller than $10^{-6}$ Ry/Bohr and the energy is converged to $10^{-8}$ Ry. The optimized cutoff energy of 70 Ry seems to suffice in giving the results with the required precision. Throughout the entirety of the calculations, a Monkhorst-Pack \cite{PhysRevB.13.5188} mesh of $30\times30\times1$ is set for the Brillouin zone integration. A large vacuum of 20 angstroms is placed between the layers in the $z$-direction of the vdW heterostructure, to keep the layers far apart and avoid any interlayer interaction. To compute the band structure, the Gaussian smearing method \cite{PhysRevB.28.5480} is exercised, and a smearing width of $\sigma=0.01$ Ry is chosen.

To compute the lattice mismatch between the two layers of graphene and HfS$_2$, one can take the mechanism below~\cite{C5EE03490F}:

\begin{equation}\label{Eq0}
\eta=(a_G-a_{HfS_2})/a_{HfS_2}.
\end{equation}

where $\eta$ is the lattice mismatch and $a_G$ and $a_{HfS_2}$ are the optimized lattice parameters of $3\times3\times1$ supercell of graphene and $2\times2\times1$ supercell of HfS$_2$, respectively. 

The interface binding energy of vertical G/HfS$_2$ contact per each atom of graphene is measured as

\begin{equation}\label{Eq1}
E_b=[E_{G/HfS_2}-(E_G+E_{HfS_2})]/N_G,
\end{equation}
where $E_b$ is the binding energy, $E_{G/HfS_2}$ is the total energy of G/HfS$_2$ heterostructure, $E_G$ and $E_{HfS_2}$ are total energies of monolayers of graphene and HfS$_2$ respectively; and $N_G$ is the number of carbon atoms present in the layer of graphene used in the construction of the heterostructure.

The averaged electron density difference normal to the interface of the heterostructure is given by 

\begin{equation}\label{Eq2}
\begin{split}
& \Delta \rho = \iint \rho_{G/HfS_2}(x,y,z)dxdy \\
& - \iint \rho_{HfS_2}(x,y,z)dxdy - \iint \rho_{G}(x,y,z)dxdy.
\end{split}
\end{equation}

The sum above contains $\rho_{G/HfS_2}(x,y,z)$, $\rho_{HfS_2}(x,y,z)$ and $\rho_{G}(x,y,z)$ which are the charge densities located at $(x,y,z)$,  pertaining to the heterostructure; and HfS$_2$ and graphene monolayers, respectively.

\begin{figure}
\vspace{2mm}
\includegraphics*[scale=0.36]{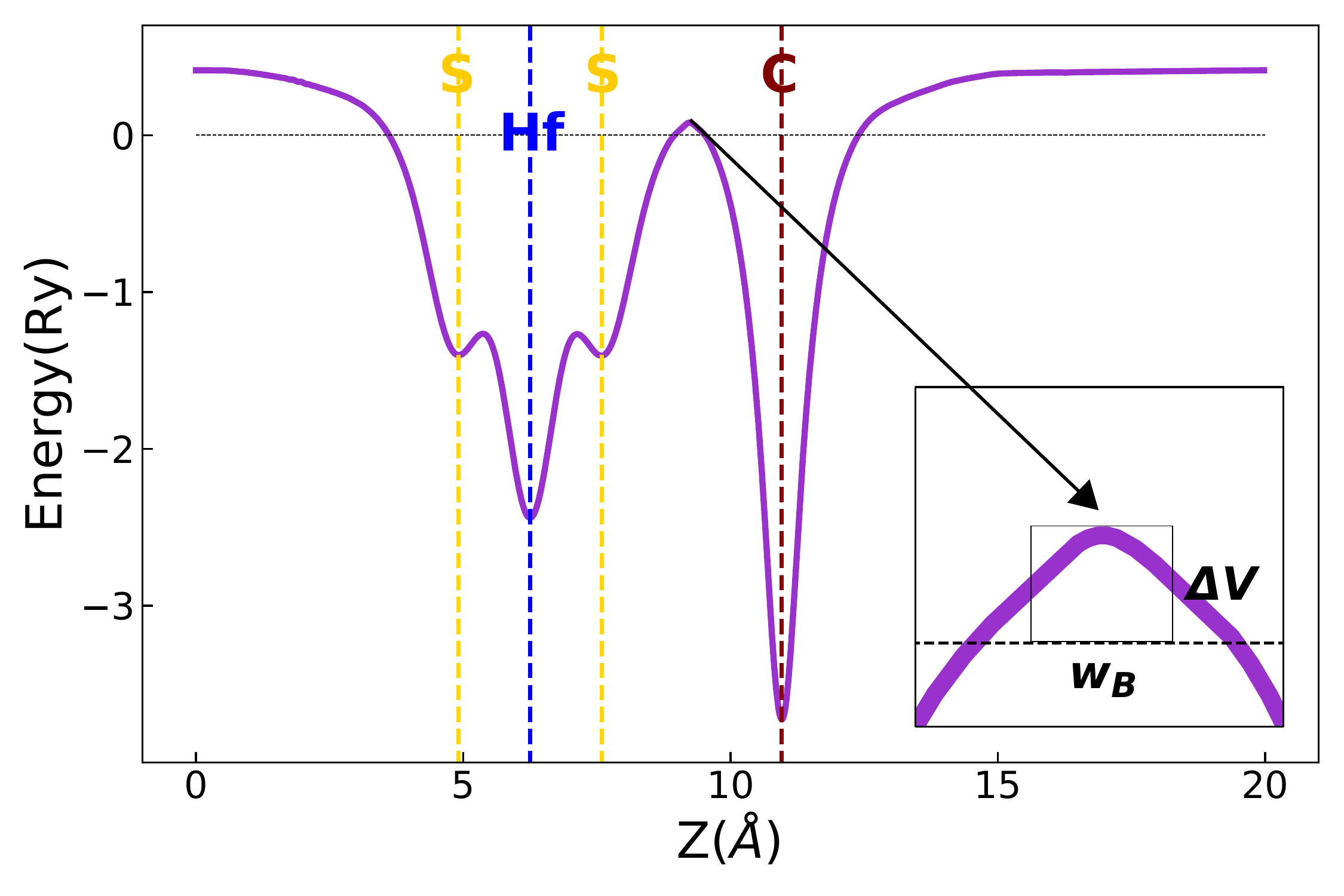}
 \caption{
 Average electrostatic potential of G/HfS$_2$ in planes normal to the interface while the distance between the layers is 3.28 angstroms and no strain is placed on the device. $\Delta V$ and $w_B$ are clearly indicated on the square potential barrier.    
}
\label{1}
\end{figure}

The tunneling barrier has to be obtained to calculate $T_B$, a factor which is apprehensible via the average electrostatic potential in planes normal to the interface of the contact. In this literature, a square potential barrier is taken to replace the real potential barrier. We can compute The maximum $T_B$ through the following formula \cite{C5NR06204G}

\begin{equation}\label{Eq3}
T_B = exp (-2\frac{\sqrt{2m\Delta V}}{h} \times w_B),
\end{equation}
where $m$ is the mass of the electron, $\Delta V$ is described as the potential energy above Fermi level between the two components of the heterostructure, $h$ is the Planck's constant and $w_B$ is the full width at half maximum of the potential energy above Fermi level ($\Delta V$). To provide clarity on how these parameters ($\Delta V$ and $w_B$) are determined, the interpolated graph of Fig. \ref{1} is drawn. As shown in this figure, the barrier height and the barrier width are the height ($\Delta V$) and width ($w_B$) of the square potential barrier. 

By changing the crystal lattice parameter, the in-plane biaxial strain is put on G/HfS$_2$ contact, and the calculation of said strain is performed through the medium of this next equation:

\begin{equation}\label{Eq4}
\epsilon=[(a-a_0)/a_0]\times 100\%.
\end{equation}

Here $a$ and $a_0$ are indicators of the lattice parameters with and without the exertion of strain modifications, respectively.

To calculate the interface dipole moment, the Helmholtz relation is applicaple \cite{PhysRevB.81.125403,jackson2007classical},

\begin{equation}\label{Eq5}
\Delta\mu=\frac{\epsilon_0 A}{e}|\Delta W_F|,
\end{equation}

where $\epsilon_0$ is the vacuum permittivity or electric constant, $A$ is the interface area, $e$ is the elementary charge and $|\Delta W_F|$ is the absolute value of work function difference between G/HfS2$_2$ contact and the free-standing semiconductor of HfS$_2$ \cite{PhysRevB.79.195425}. 

\section{RESULTS AND DISCUSSION}\label{III}

\begin{figure*}
\vspace{2mm}
\includegraphics*[scale=0.054]{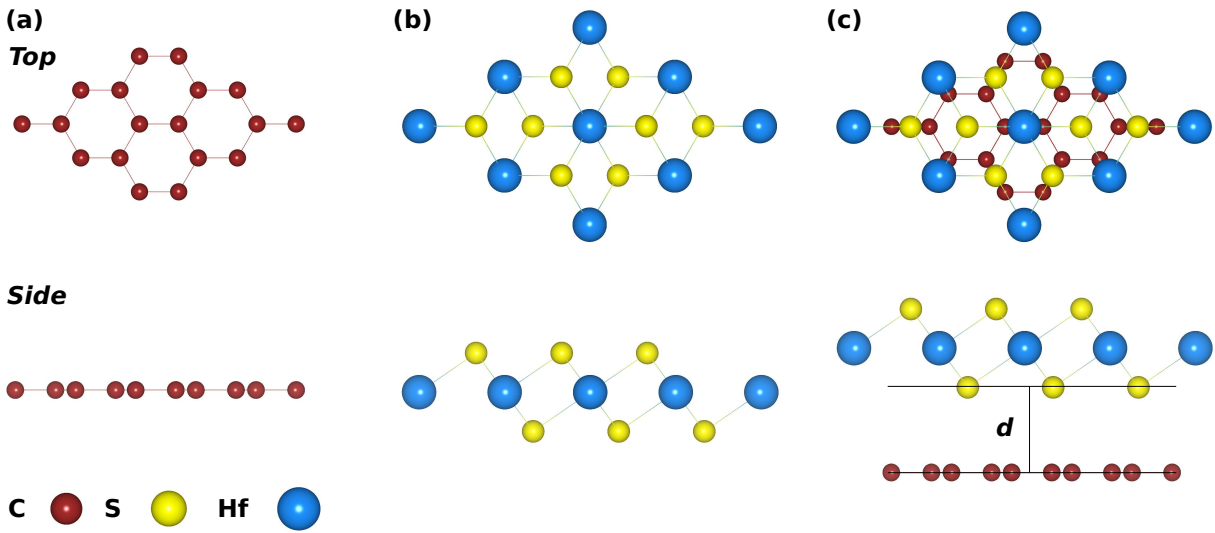}
\vspace{2mm}
 \caption{
 Schematic graphics of optimized crystalline structures of (a) $3\times3\times1$ supercell of graphene, (b) $2\times2\times1$ supercell of HfS$_2$ and (c) G/HfS$_2$ van der Waals heterostructure. 
}
\label{2}
\end{figure*}

Before diving into the world of heterostructures, each monolayer is optimized and inspected fully to reach the criteria of force and energy mentioned in the previous section. This optimization shows that the lattice constant of hafnium disulphide is 3.64 angstroms, agreeing to prior studies \cite{YUAN2018919,ZHAO2016302};. The lattice constant of graphene is calculated to be 2.47 angstroms, same as past reports~\cite{Wang2019,WANG2019778}. After the thorough vetting of these monolayers in their unit cells is performed, the apparent lattice mismatch between them has to be dealt with; the reason for this being the fact that the contact is not desirable if constructed with a substantial percentage of mismatch. The graphene unit cell is turned into a $3\times3\times1$ supercell and the monolayer of HfS$_2$ is transformed into a structure of $2\times2\times1$ times its original unit cell. The changes applied to give the new lattice mismatch to be approximately 1.58\%, computed through Eq. \ref{Eq0}; an amount more than suitable for further simulations. A mismatch, however small, has to be compensated for when stacking these two monolayers on top of each other. To atone for this discrepancy, one of the layers has to undergo slight changes in its lattice structure by the employment of tensile or compressive biaxial strain, the mechanism of which goes according to Eq. \ref{Eq4}. Due to the sensitive nature of structural properties possessed by HfS$_2$ monolayer, causing it to change drastically under tiny amounts of strain \cite{WU20171}, graphene supercell is the sole receiver of compression. At the same time, the layer of HfS$_2$ remains fixed. This compressed supercell of graphene is depicted in Fig. \ref{2}(a) while the optimized supercell of HfS$_2$, untouched by any modification, is shown in part (b). 

\begin{figure}[!htb]
\includegraphics*[scale=0.40]{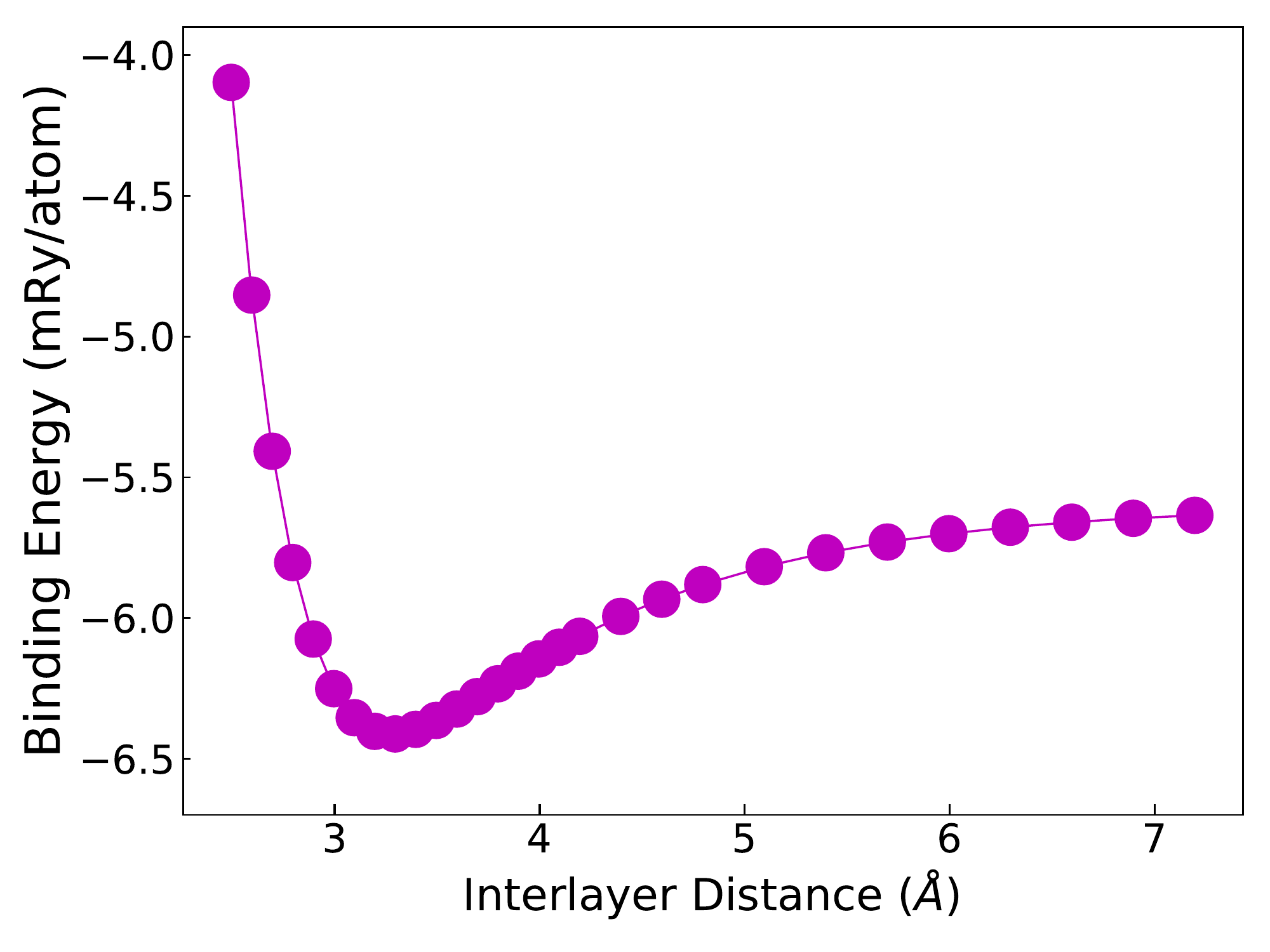}
 \caption{
 Binding energy of G/HfS$_2$ contact drawn against the interlayer distance.  
}
\label{3}
\end{figure}

Once the mismatch is overcome, to form the most stable contact, the choice of stacking configuration comes into focus. As it was mentioned before, HfS$_2$ is immensely similar to ZrS$_2$ in solid-state qualities. Suppose zirconium disulphide takes on a specific stacking configuration as the most stable when put on top of graphene. In that case, there can be no doubt that the same arrangement is proper regarding hafnium disulphide. Wang et al. \cite{WANG2019778} recently accomplished this, and we choose the most stable stacking configuration among the three most usual stacking formations. The same arrangement is selected in this article and is displayed in Fig. \ref{2}(c). 

\begin{figure*}[!htb]
\vspace{2mm}
\includegraphics*[scale=0.28]{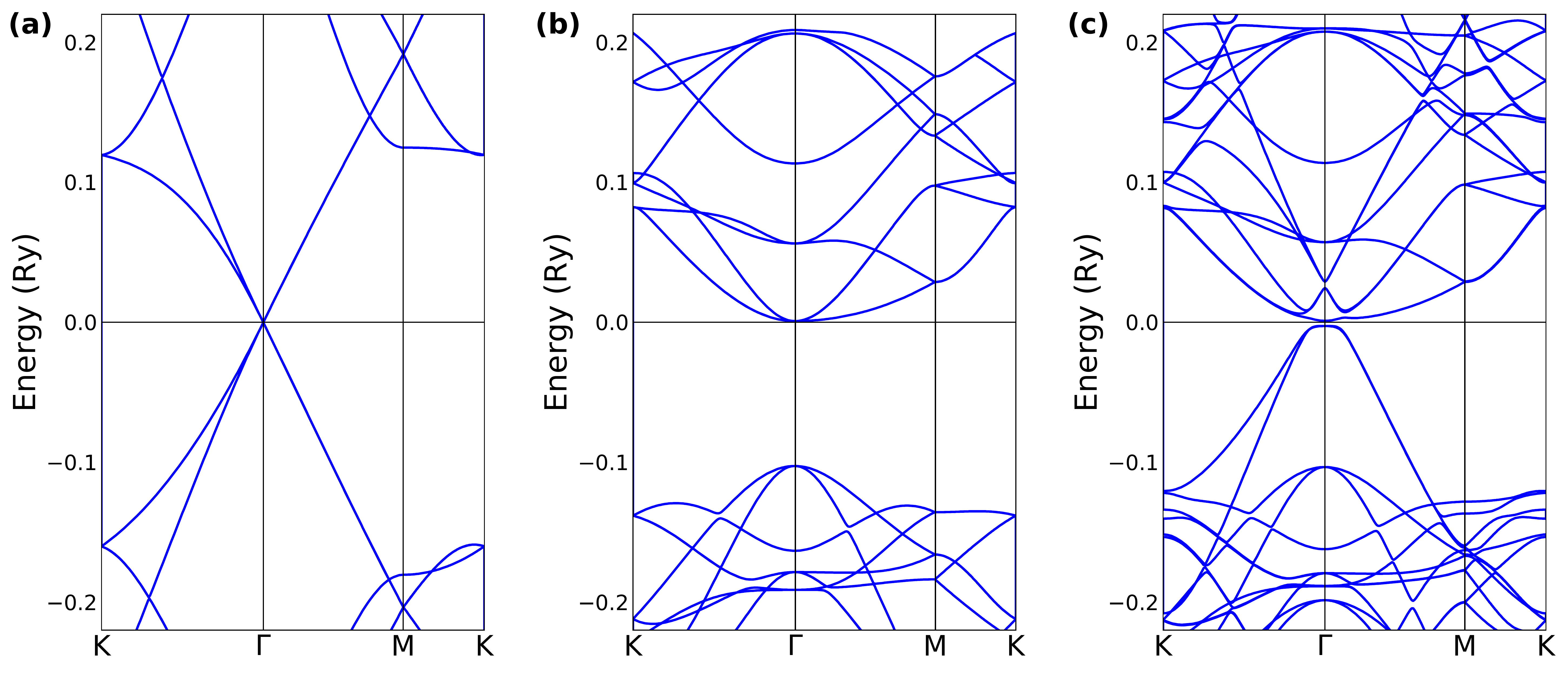}
 \caption{
 Electronic band structures of (a) graphene with a $3\times3\times1$ supercell, (b) HfS$_2$ with a $2\times2\times1$ supercell and (c) heterostructure of G/HfS$_2$. It should be noted that this band structure of G/HfS$_2$ contact is computed at the interlayer distance of 3.28 angstroms and the energy value of zero corresponds to the Fermi level. 
}
\label{4}
\end{figure*}

\begin{figure}[!htb]
\vspace{2mm}
\includegraphics*[scale=0.32]{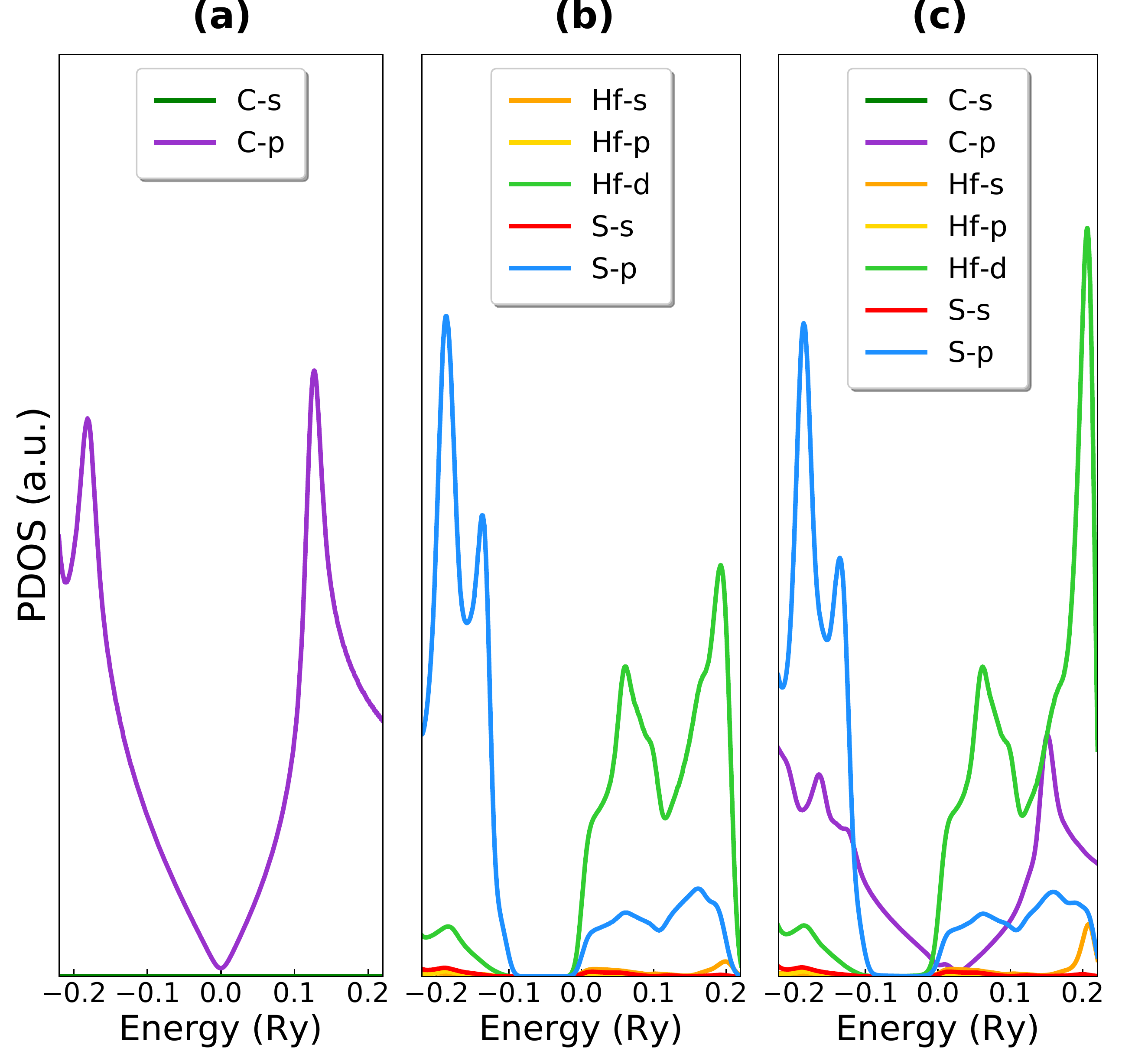}
 \caption{
 Partial Density of States of (a) graphene with a $3\times3\times1$ supercell, (b) HfS$_2$ with a $2\times2\times1$ supercell and (c) heterostructure of G/HfS$_2$ at its most stable form. The enegy range of the figure is the same as the one in Fig. \ref{4} and a.u. stands for arbitrary units. 
}
\label{5}
\end{figure}

After deciding on a stacking alignment, for the heterostructure to reach maximum stability, there is one more factor that needs to be considered, and that is the interlayer distance. The finding energy (Eq. \ref{Eq1}) is calculated for different intervals between the two supercells and is plotted in Fig. \ref{3}. It is easily understood from this graph that the interlayer distance of 3.28 angstroms provides the least binding energy and in turn, the most stable contact of graphene and HfS$_2$. The binding energy in correspondence with this most stable separation stretch is -6.41 mRy and the atoms included in the precise measurement of the interlayer distance are shown in Fig. \ref{2}(c).

 Electronic band structures of pristine $3\times3\times1$ supercell of graphene and $2\times2\times1$ supercell of HfS$_2$ are evaluated and exhibited in Fig. \ref{4}(a) and (b), respectively. As can be seen in Fig. \ref{4}(a), the Dirac cone of graphene is very well intact across the Fermi level but is positioned at the $\Gamma$-point; contrary to the primitive cell of this two-dimensional monolayer where the Dirac cone is mapped to the $K$-point. $K$ and $K^\prime$ points are folded to the $\Gamma$-point in this $3\times3\times1$ supercell of graphene, causing the Dirac cone to change its whereabouts \cite{doi:10.1063/1.4807738}. The primitive cell of HfS$_2$, much like the monolayer of ZrS$_2$, owns a bandgap that is indirect \cite{doi:10.1002/pssb.201700033}, but this material is transformed into a $2\times2\times1$ supercell; its band structure becomes different in a way that could be seen in Fig. \ref{4}(b). This transformation changes the indirect bandgap of this substance into a direct bandgap, located at the $\Gamma$-point. The same development occurs in the single layer of ZrS$_2$ when faced with similar transformation \cite{C5EE03490F,WANG2019778} and due to the likeness between the two materials, we can establish this happening as expected and normal. This direct bandgap of HfS$_2$ pristine supercell has the value of 1.40 eV (0.1032 Ry). At the time of writing this article and based on the knowledge gathered to the best of author's capabilities, there exists only one more study that has estimated the bandgap of $2\times2\times1$ supercell of pristine HfS$_2$ using PBE functional, giving it at 1.45 eV \cite{kingori2020twodimensional}. One can accept this amount as a rough match with the result procured in this work. 
  
\begin{table*}[!htb]
\caption{\label{geometry}
Schottky barrier height (SBH), tunneling barrier height ($\Delta V$) and tunneling probability ($T_B$) of G/HfS$_2$ contact under different interface separations
}
\vspace{2mm}
 \begin{tabular*}{\textwidth}{c @{\extracolsep{\fill}} cccccc}
Interlayer Distance ($\AA$) & 2.80   & 3.05  & 3.28  & 3.50  & 3.75  & 4.00 \\
\noalign{\vskip 0.2mm}
\hline
\hline
\noalign{\vskip 1mm}
SBH(eV)  &0.00 &  0.00 & 0.00 & 0.00 &0.00 &0.00 \\
$\Delta V$(eV) &0.00 &0.10 &1.09 &1.88 &2.65 &3.30  \\
$T_B$ &100.00\% &  99.74\% & 95.33\% & 91.30\% &86.85\% &82.08\%  \\
\noalign{\vskip 0.6mm}

\hline
\hline
\end{tabular*}
\label{T1}
\end{table*}

\begin{figure*}[!htb]
\vspace{2mm}
\includegraphics*[scale=0.21]{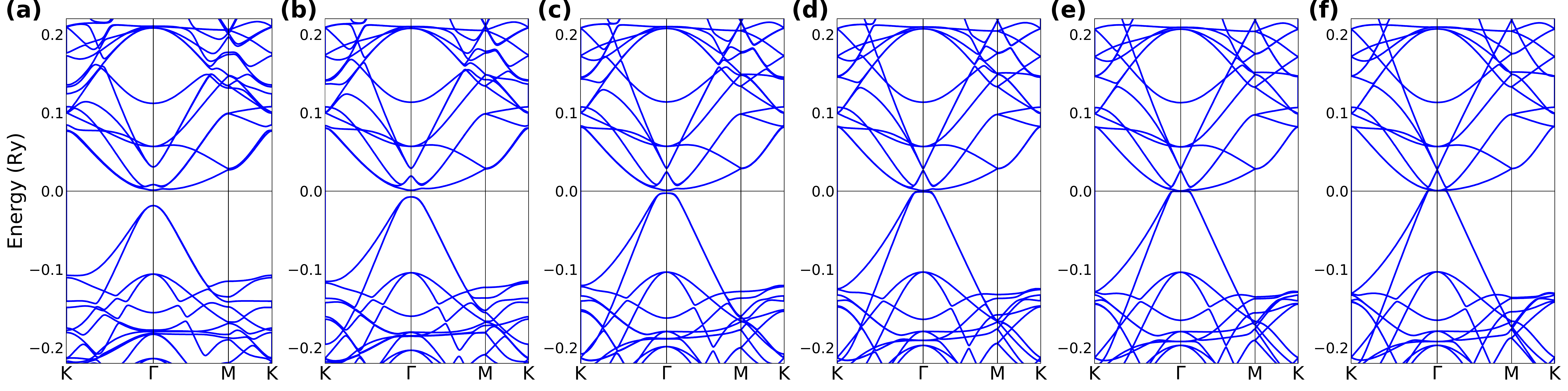}
 \caption{
 Electronic band structures of G/HfS$_2$ vdW heterostructures with different interlayer distances of (a) 2.80$\AA$ (b) 3.05$\AA$ (c) 3.28$\AA$ (d) 3.50$\AA$ (e) 3.75$\AA$ (f) 4.00$\AA$. The energy value of zero corresponds to the Fermi level and is pinned to the CBM of HfS$_2$.
 }
\label{6}
\end{figure*}

\begin{figure*}[!htb]
\vspace{2mm}
\includegraphics*[scale=0.35]{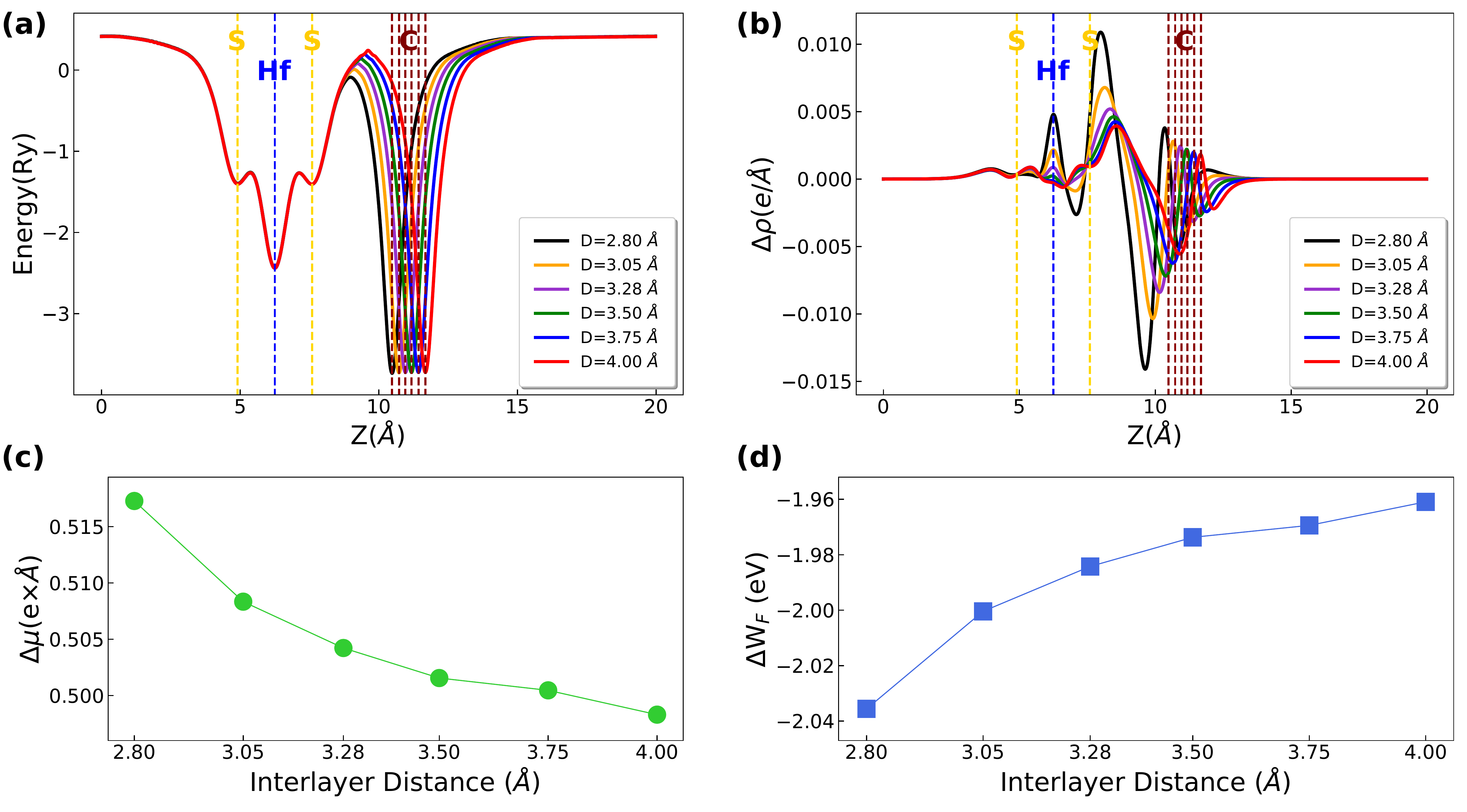}
 \caption{
 (a) Average electrostatic potential in planes normal to the G/HfS$_2$ contact, (b) averaged electron density difference normal to the interface of G/HfS$_2$, (c) interface dipole moment of G/HfS$_2$ and (d) work function difference of G/HfS$_2$ and pristine HfS$_2$; under different interlayer distances of 2.80$\AA$, 3.05$\AA$, 3.28$\AA$, 3.50$\AA$, 3.75$\AA$ and 4.00$\AA$.
 }
\label{7}
\end{figure*}
  
Part (c) of Fig. \ref{4} portrays what can be called the band structure of the most stable heterostructure of graphene and HfS$_2$ in terms of stacking configuration and interlayer distance. As this shape clearly shows, this band structure is very nearly a simple superposition of the two pristine band structures of graphene and HfS$_2$ shown in Fig. \ref{4}(a) and (b), respectively. The interactions between the two layers are indeed weak vdW interactions, preserving the electronic structures of both graphene and HfS$_2$. There are, however, slight changes in this band structure of G/HfS$_2$ contact when compared with the pristine band structures of each component. There is an opening in the Dirac cone of graphene, and a bandgap of 0.06 eV appears in the band structure.
In comparison, the distance between the conduction band minimum (CBM) and the valence band maximum (VBM) of HfS$_2$ is more considerable by an approximate amount of 0.01 eV. Based on the actuality that the Fermi level of the system intersects with the CBM of HfS$_2$ and the bandgap of the semiconductor stays empty of newly created states.  The contact nature is $n$-type Ohmic, and hence, the Schottky barrier height (SBH) of this structure is zero \cite{ref1}. 

To specify the orbitals that contribute to the band structures drawn in Fig. \ref{4}, the partial density of states (PDOS) is plotted in Fig. \ref{5}. This figure shows the PDOS of pristine graphene and HfS$_2$ supercells in parts (a) and (b), clearly depicting the fact that carbon's 2p orbital, as well as hafnium's 5d and sulfur's 3p, are the most effective in the creation of the associated band structures in the determined energy range. Part (c) of Fig. \ref{5}, displaying a partial DOS of G/HfS$_2$ heterostructure, further approves of the notion that no new states are added within the bandgap of HfS$_2$ and the bands of pristine supercells of graphene and HfS$_2$ remain preserved near the bandgap edges. 
 An overlap occurs between C-2p and Hf-5d orbitals at the CBM of HfS$_2$ around the $\Gamma$-point (See Fig. \ref{4} and \ref{5} ).
 
The interlayer distance of the heterostructure is one of the factors that have a high probability of change during the construction process of G/HfS$_2$ contact. Previous studies of graphene-based heterostructures have also shown that changing the amount of the interface space can transform a Schottky contact to an Ohmic one or vice versa \cite{Wang2019,SLASSI2020154800}. Hence, investigating this structure as it goes through a change in its interlayer distance is only mandatory. The different selected lengths of 2.80, 3.05, 3.28, 3.50, 3.75 and 4.00 angstroms are created in the contact wich explore the effect of the interlayer gap in this system. The band structures of these compositions are shown in Fig. \ref{6}. In parts (b) through (f) of this figure, the CBM of HfS$_2$ is pinned to the Fermi level which holds the energy value of zero and the system remains $n$-type Ohmic throughout this change in its interlayer distance. As shown in Table \ref{T1}, SBH does not move from the amount of zero.
The interlayer distance is smaller than 3.28 $\AA$; a subtle shift happens to the band of graphene and the valence band of HfS$_2$.
As the distance decreases, the bandgap of HfS2$_2$ rises to amounts of 1.43 eV and 1.46 eV. The overlap between the orbitals of C-2p and Hf-5d is also affected, showing an increase in the interface interaction. To go further in decreasing the interlayer distance means to get energetically distant from the most stable structure (Fig. \ref{3}). This results in a very distorted band structure which is no longer preservative of the heterostructure's electronic properties. One can notice the beginning of such action in Fig. \ref{6}(a). Fig. \ref{6} (d), (e) and (f) show the band structures of the heterostructures with interlayer distances of more than 3.28 $\AA$. As the distance increases, the valence band of HfS$_2$ and the band of graphene shift up while the CBM of HfS$_2$ remains located at the Fermi level, keeping the contact's $n$-type Ohmic identity intact. As a result, the bandgap of HfS$_2$ sees a minimal decrease through this change. The interface interaction is also lowered, a consequence of the reduced overlap between the orbitals of C-2p and Hf-5d. Upon further increase of the interlayer distance to amounts of 4.25 $\AA$, 4.50$\AA$, 4.75 $\AA$ and 5.00 $\AA$, the contact's band structures continue to show very subtle shifts in both bands of graphene and HfS$_2$ following the same pattern as before. These small changes do not move the Fermi level from its position at the CBM of HfS$_2$, preserving the system as $n$-type Ohmic.   

Carrier injection is directly affected by SBH, but this barrier is not the sole contributor to electron injection efficiency. Other factors that are valuable in determining this efficiency are the tunneling barrier ($\Delta$V) and the tunneling probability ($T_B$). $\Delta$V and $T_B$ are calculated based on Eq. \ref{Eq3} and Fig. \ref{1} and are listed in Table \ref{T1}. $T_B$ is on a monotonous incline as the distance between the layers drops from 4.00 $\AA$ to 2.80 $\AA$, suggesting a higher injection efficiency as the layers come closer to each other. The reason for this is closely related to $\Delta$V, and as the barrier height sees a decrease, the percentage of efficiency experiences a substantial growth (depicted in Fig. \ref{7}(a)).

As the interface distance gets smaller, the chance of the layers' electronic wavefunctions extending into one another, causing different degrees of orbital overlap and band hybridization, becomes greater. Through charge transfer between layers of HfS$_2$ and graphene. Thus, the averaged electron density difference ($\Delta\rho$) of G/HfS$_2$ contact is calculated based on Eq. \ref{Eq2} and is shown in Fig. \ref{7}(b). In this figure, positive numbers are an indicator of charge accumulation at that area while the negative amounts depict charge depletion. The most disturbing area is positioned in between the layers, and there is even a small redistribution of charge for Hf and S atoms, revealing the strong interaction of the two layers. An increase follows the overall shape that the transfer of the electrons takes place from graphene to HfS$_2$ and the decline of the interlayer distance in $\Delta\rho$. According to Bader charge analysis performed using Yu-Trinkle integration \cite{doi:10.1063/1.3553716,OTERODELAROZA20141007,OTERODELAROZA2009157}, more electrons transfer from graphene to HfS$_2$ with the decrease of interlayer distance (0.0757, 0.0766, 0.0774, 0.0785, 0.0810 and 0.0890 $|e|$ for interlayer distances of 4.00, 3.75, 3.50, 3.28, 3.05 and 2.80 $\AA$). This analysis is in complete agreement with what is deduced from Fig. \ref{7}(b) and a lower interface distance can be expected to have a positive effect on the electron injection efficiency of G/HfS$_2$ heterostructure. 

\begin{table*}[!htb]
\caption{\label{geometry}
Interlayer distance, Schottky barrier height (SBH), tunneling barrier height ($\Delta V$) and tunneling probability ($T_B$) of G/HfS$_2$ contact under different amounts of strain.
}
\vspace{2mm}
 \begin{tabular*}{\textwidth}{c @{\extracolsep{\fill}} cccccc}
Strain & -4\%   & -2\%  & 0\%  & 2\%  & 4\% & 6\% \\
\noalign{\vskip 0.2mm}
\hline
\hline
\noalign{\vskip 1mm}
Interlayer Distance ($\AA$) & 3.21   & 3.24   & 3.28  & 3.31  & 3.35  & 3.38 \\
SBH(eV)  &0.00 &  0.00 & 0.00 & 0.00 &0.00 &0.19 \\
$\Delta V$(eV) &1.14 &1.11 &1.09 &1.05 &1.02 &0.98  \\
$T_B$ &95.15\% &  95.24\% & 95.33\% & 95.46\% &95.58\% &95.69\%  \\
\noalign{\vskip 0.6mm}

\hline
\hline
\end{tabular*}
\label{T2}
\end{table*}

\begin{figure*}[!htb]
\vspace{2mm}
\includegraphics*[scale=0.21]{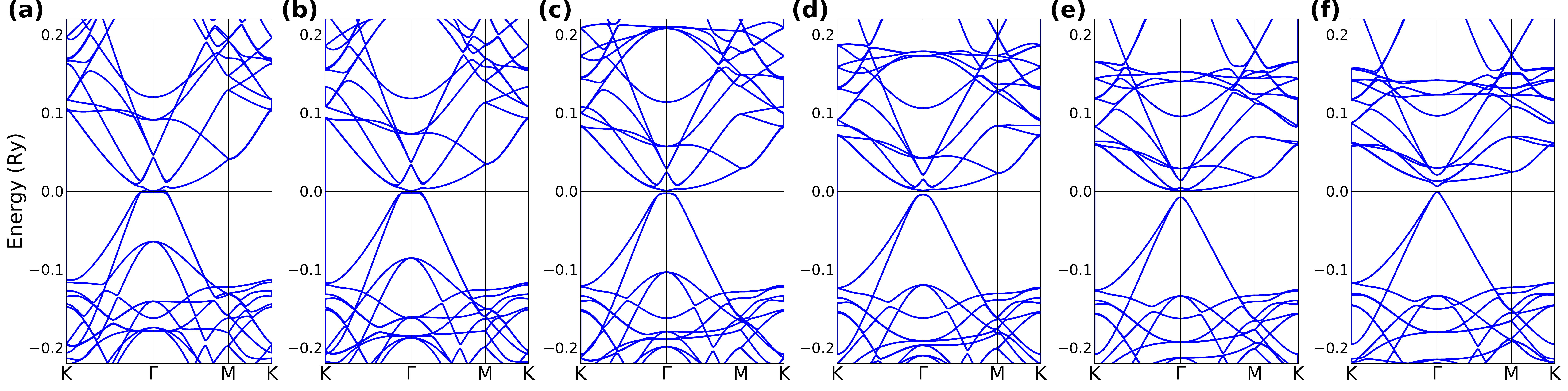}
 \caption{
 Electronic band structures of G/HfS$_2$ vdW heterostructures under different amounts of strain, (a) -4\% (b) -2\% (c) 0\% (d) 2\% (e) 4\% (f) 6\%. The Fermi level is set to zero.
 }
\label{8}
\end{figure*}

\begin{figure*}[!htb]
\vspace{2mm}
\includegraphics*[scale=0.35]{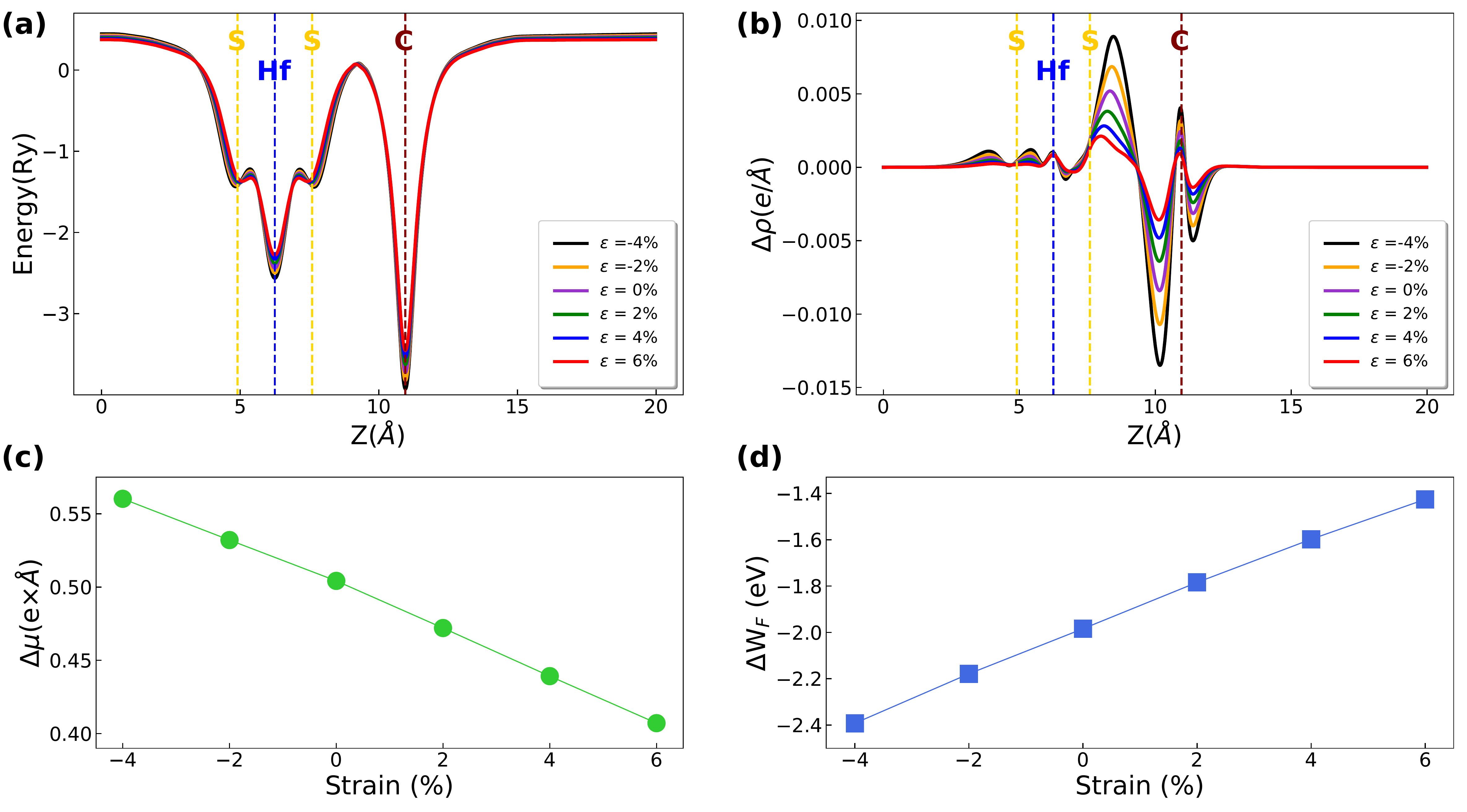}
 \caption{
 (a) Average electrostatic potential in planes normal to the G/HfS$_2$ contact, (b) averaged electron density difference normal to the interface of G/HfS$_2$, (c) interface dipole moment of G/HfS$_2$ and (d) work function difference of G/HfS$_2$ and pristine HfS$_2$; under different amounts of strain, -4\%, -2\%, 0\%, 2\%, 4\%, 6\%.
 }
\label{9}
\end{figure*}

As HfS$_2$ monolayer comes into contact with graphene, a transfer of free electrons takes place between the two subjects. This phenomenon can cause an interface dipole moment ($\Delta\mu$) and a built-in electric field, resulting in interlayer interaction \cite{PhysRevB.90.085415}. $\Delta\mu$ determine the effect of interlayer distance on orbital overlap more vividly. In Fig. \ref{7}(c), $\Delta\mu$ is drawn against different interlayer distances. As the interface distance increases, $\Delta\mu$ goes on a decline, confirming the fact that a lower interlayer distance is indeed influential in intensifying the interlayer overlap, resulting in the achievement of the low resistance contact. This charge transfer between the layers also possesses the power to give rise to band bending via aligning the Fermi level difference between graphene and HfS$_2$. This bending is evaluated by the work function difference ($\Delta$W$_F$) between the G/HfS$_2$ contact and the free-standing semiconductor of HfS$_2$ and is presented in Fig. \ref{7}(d). As is seen, the absolute value of $\Delta$W$_F$ is on a steady decrease due to the weakened charge transfer and interlayer interaction. The sign of this quantity determines the direction in which the bands bend. If, as in this case, $\Delta$W$_F$ is smaller than zero, the bands turn downward, indicating an $n$-type doping in the layer of HfS$_2$.

Building the heterojunction of G/HfS$_2$ in real life is usually performed with the help of a substrate. Among the many effects that this substrate can have on the heterostructure, the strain is one that cannot be ignored. Earlier work suggests strain as an instrument for changing an Ohmic contact to one of Schottky nature, with the other way around also a possibility \cite{WANG2019778,C8CP03508C}. To see how G/HfS$_2$ changes when put under strain, the biaxial strain is applied to the system based on Eq. \ref{Eq4}. The amounts of biaxial strain studied in this work include -4\%, -2\%, 0\%, 2\%, 4\% and 6\%; their relative band structures calculated and shown in Fig. \ref{8}. In parts (a) and (b) of this figure, where the applied strain is compressive, the CBM of HfS$_2$ remains fixed on the Fermi level (specifying no change in the system's $n$-type Ohmic status), while its VBM moves up with the increase in compressive strain. The band of graphene is also shifted upward when going through this process. Due to this change, the bandgap of HfS$_2$ is narrowed down to 1.17 eV and 0.88 eV for the strain amounts of -2\% and -4\%, respectively. The increased overlap of C-2p and Hf-5d is also very evident, pointing out the fact that compressive strain has the capability of escalating the interlayer interaction. Fig. \ref{8}(d), (e) and (f) show the band structures of G/HfS$_2$ contacts when subjected to tensile strain. The weakened band hybridization makes for a downward rearrangement of HfS$_2$ valence band and the band of graphene relative to the Fermi level, increasing the bandgap of HfS$_2$ monolayer to 1.64 eV, 1.83 eV and 2.00 eV for 2\%, 4\% and 6\% of tensile strain, respectively. Although the bandgap of HfS$_2$ becomes indirect for the tensile strain of 4\% and 6\%, the bandgap reported here is still of the distance between the CBM of HfS$_2$ and the highest spot of the valence band positioned in $\Gamma$-point. This is to show the reduced overlap as the tensile strain increases. The heterojunction remains $n$-type Ohmic through 2\% and 4\% of strain and a Schottky barrier only appears when the substantial tensile strain of 6\% is applied. As a consequence of this, the CBM of HfS$_2$ is no longer attached to the Fermi level. In calculating SBH, the Schottky-Mott rule \cite{ref1} is employed, as there are no new electronic states formed in the bandgap of HfS$_2$ (denoting the absence of Fermi level pinning in the contact). Based on this regulation, SBH is evaluated by the difference of the contact's Fermi level and the band edges of the semiconductor. The energy gap between the CBM of HfS$_2$ and the Fermi level is refered to as the $n$-type SBH ($\Phi_{B,n}^0$), while the $p$-type SBH ($\Phi_{B,p}^0$) is the energy variance between the Fermi level and the VBM of the semiconductor. The sum of these two barriers ($\Phi_{B,n}^0$+$\Phi_{B,p}^0$) is roughly equal to the bandgap of HfS$_2$. If the $p$-type SBH is larger than the $n$-type SBH ($\Phi_{B,n}^0<\Phi_{B,p}^0$), the vdW heterostructure is addressed as an $n$-type Schottky contact. This is true for the G/HfS$_2$ heterojunction when subjected to a 6\% tensile strain. Appling biaxial strain  from 6\% to -4\%, the distance between the layers changes as well as the atoms of sulfur are forced to reorganize and settle in a closer position to the monolayer of graphene. Computed amounts of SBH and interlayer distance are summarized in Table \ref{T2}.

\begin{figure*}[!htb]
\includegraphics*[scale=0.23]{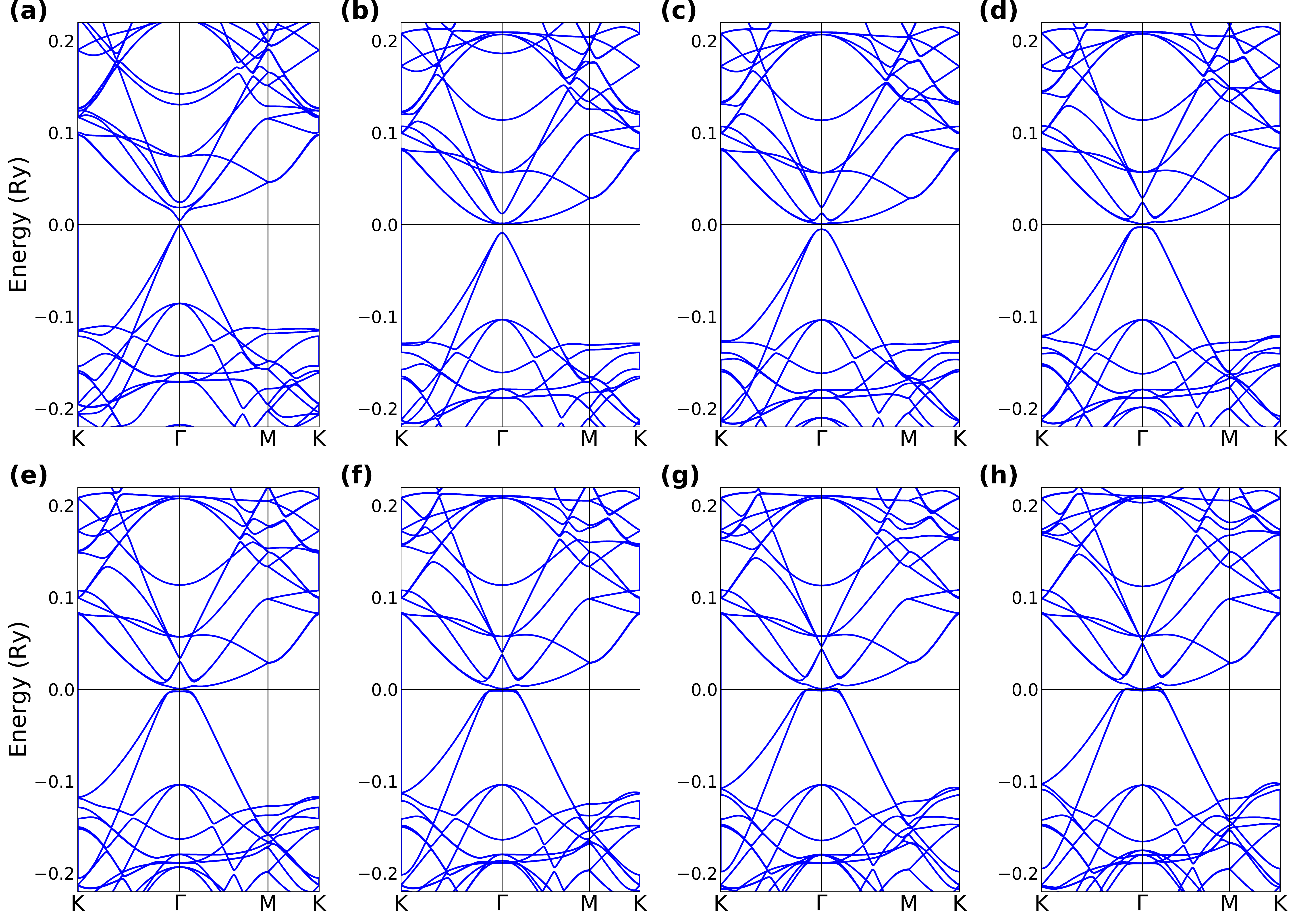}
\caption{
Electronic band structures of G/HfS$_2$ vdW heterostructures subjected to different perpendicular electric fields in the amounts of (a) -0.6 V/$\AA$, (b) -0.4 V/$\AA$, (c) -0.2 V/$\AA$, (d) 0.0 V/$\AA$, (e) 0.2 V/$\AA$, (f) 0.4 V/$\AA$, (g) 0.6 V/$\AA$ and (h) 0.8 V/$\AA$.
}
\label{10}
\end{figure*}

In the event of an increase in the amount of strain from -4\% to 6\%, the tunneling barrier ($\Delta$V) goes in the opposite direction, experiencing a very subtle decline. On this account, the tunneling probability ($T_B$) sees an increase in the same magnitude. The alterations made to the quantities of $\Delta$V and $T_B$; however, are not of considerable size as $T_B$ stays almost equal to 95\% under different numbers of strain (Table \ref{T2}). The average electrostatic potential in planes normal to the interface of G/HfS$_2$ (Fig. \ref{9}(a)), where $\Delta$V and $w_B$ are nearly unchanged throughout the cycle of tensile and compressive strain. The charge transfer between the layers of G/HfS$_2$, offering a better picture of how the orbitals overlap and the bands mix, is portrayed through the computation of $\Delta\rho$ in Fig. \ref{9}(b). As mentioned before, the positive and negative quantities in the figure specify charge accumulation and depletion, respectively. It is transparent in Fig. \ref{9}(b) that higher amounts of compressive strain give a larger charge transfer weight in comparison with tensile strain, resulting in a higher orbital overlap when moving from the strain amount of 6\% to -4\%. Charge transfer simulations performed by Bader charge population analysis also show that the transfer of electrons from the monolayer of graphene to HfS$_2$ increase as tensile strain turns into compressive strain (0.0234, 0.0384, 0.0573, 0.0785, 0.1006 and 0.1248 $|e|$ for strain amounts of 6\%, 4\%, 2\%, 0\%, -2\% and -4\%). The larger charge transfer rates in the application of compressive strain promise higher injection efficiencies in the heterojunction of G/HfS$_2$.

Fig. \ref{9}(c) depicts $\Delta\mu$ as a function of different amounts of strain. When the applied biaxial strain goes from 6\% to -4\%, $\Delta\mu$ sees a steady incline, suggesting a higher interlayer interaction and orbital overlap. The same routine happens to the absolute value of $\Delta$W$_F$ (Fig. \ref{9}(d)) and the Fermi level alignment between HfS$_2$ and monolayer of graphene causes an increase in band bending. These figures go along perfectly with what was already seen in Fig. \ref{8} and Fig. \ref{9}(b) in terms of band hybridization. The sign of $\Delta$W$_F$ indicates that band bending stays downward throughout the different quantities of strain, keeping the doping in the monolayer of HfS$_2$ $n$-type. 

Applying an external electric field in a direction normal to the surface of two-dimensional heterostructures has proven to be a useful method for tuning their electronic properties \cite{C7CP01852E,C8CP02190B}. Based on this, the most stable heterojunction of G/HfS$_2$ with the layers separated at 3.28 $\AA$, is subjected to different amounts of the electric field in the $z$-direction. These electric fields include -0.6, -0.4, -0.2, 0.0, 0.2, 0.4, 0.6 and 0.8 V/$\AA$. The band structures in correspondence to these quantities are calculated and drawn in Fig. \ref{10}. Parts (a), (b) and (c) of this figure show the bands as they change due to the negative electric field. It is obvious that with the increase in this field, the Dirac cone of graphene moves towards the CBM of HfS$_2$, passing it with the field at -0.6 V/$\AA$. This causes the $n$-type Ohmic contact to change to an $n$-type Schottky heterostructure, with the SBH amount of 0.26 eV. Fig. \ref{10}(e) through (h) show the electronic band structures of G/HfS$_2$ when put through different quantities of positive electric field. It is seen that graphene's Dirac cone moves further away from the CBM of HfS$_2$ as the field goes on an incline. This signifies that from the negative electric field of -0.4 V/$\AA$ to the positive field of 0.8 V/$\AA$, the heterojunction of G/HfS$_2$ remains $n$-type Ohmic.

Based on what is given earlier, graphene can act as both gate electrodes and source or drain in the structure of a transistor. Proposing in this article, acting as the channel, is the monolayer of HfS$_2$; with the addition of the already mentioned native high-$k$ dielectric, HfO$_2$. 
G/HfS$_2$ contact can also be used in the construction of a two-dimensional $p$-$i$-$n$ junction. This idea originates from what is suggested in Ref. \cite{WANG2019778}, where ZrS$_2$ is used in the design of ScS$_2$/ZrS$_2$/G junction. Since graphene induces $n$-type doping in HfS$_2$, and the heterostructure of G/HfS$_2$ carries on being an Ohmic contact with high $T_B$ under different amounts of interlayer distance, strain and external electric field, its presence in a $p$-$i$-$n$ junction can be very fruitful. It broadens the choices for the other two-dimensional substance which needs to induce $p$-type doping in HfS$_2$. As mentioned in Ref. \cite{WANG2019778}, to compensate for the possible insufficient charge transfer of the contacts in the junction, an external gate voltage can be implemented to perform electrostatic doping \cite{doi:10.1063/1.1769595}. This can provide a device operable as a $p$-$n$($n$-$p$) junction or an $n$-type ($p$-type) resistor. As one example, gate tunable black phosphorus-monolayer MoS$_2$ two-dimensional $p$-$n$ junction has been experimentally created, showing great promise in broad-band photodetection and solar energy harvesting \cite{Deng2014}.

\section{CONCLUSION}\label{IV}

In short, the electronic structure of G/HfS$_2$ vdW heterostructure is investigated employing first-principles calculations. The most stable form of this contact is reached through a particular stacking alignment, and the interlayer distance of 3.28 $\AA$. The band structure of this contact at the equilibrium distance shows that the electronic properties of both layers are just about preserved, indicating the weak vdW interactions between the two materials. This heterojunction forms an $n$-type Ohmic contact which stays intact upon the change in its interlayer distance, ranging from 2.80 $\AA$ to 5.00 $\AA$, and $T_B$ reaches as high as 100\%. The compressive and tensile biaxial strain is also applied in the amounts of -4\%, -2\%, 2\%, 4\% and 6\%. The heterojunction only transforms to an $n$-type Schottky contact when subjected to a tensile strain of 6\%, showing an SBH of 0.19 eV. $T_B$ stays almost at 95\% throughout the different quantities of applied strain. Other variations of interlayer distance and strain are entirely instrumental in modulating band hybridization, orbital overlap and interface dipole moment. Tuning the position of the band structure of graphene relative to that of HfS$_2$ is proven to be possible by an external electric field perpendicular to the surface of the contact. This allows for transforming the $n$-type Ohmic system to an $n$-type Schottky one when the applied negative electric field reaches -0.6 V/$\AA$. G/HfS$_2$ can also be a part of a low resistance junction in future endeavours, with the help of electrostatic doping in case of charge transfer deficits and to boost the junction's multipotentiality. The features introduced and the results procured here are fundamental in the design and manufacture of nanodevices, while enhancing the future applications of G/HfS$_2$ contact.     

\section{ACKNOWLEDGMENTS}
Sheikh Bahaei National High Performance Computing Center (SBNHPCC) is gratefully acknowledged as the provider of needed computing facilities. SBNHPCC is supported by the scientific and technological department of presidential office and Isfahan University of Technology (IUT). The guidelines provided by Dr. S. J. Hashemifar and Dr. M. Amirabbasi are highly valued and appreciated.  

\bibliography{G-HfS2}
\end{document}